\documentclass{PoS}
\usepackage{amsmath,amssymb}
\usepackage[caption=false]{subfig}

\title{Conjugate Directions in Lattice Landau and Coulomb Gauge Fixing}

\ShortTitle{Conjugate Directions in Lattice Gauge Fixing}

\author{\speaker{R.J. Hudspith}\\
        Department of Physics and Astronomy, York University, \\
        Toronto, Ontario, M3J 1P3, Canada\\
        E-mail: \email{renwick.james.hudspith@gmail.com}}

\abstract{We provide details expanding on our implementation of a non-linear conjugate gradient method with Fourier acceleration for lattice Landau and Coulomb gauge fixing. We find clear improvement over the Fourier accelerated steepest descent method, with the average time taken for the algorithm to converge to a fixed, high accuracy, being reduced by a factor of 2 to 4. We show such improvement for the logarithmic definition of the gauge fields here, having already shown this to be the case for a more common definition. We also discuss the implementation of an optimal Fourier accelerated steepest descent method.}

\FullConference{The 32nd International Symposium on Lattice Field Theory,\\
		23-28 June, 2014\\
		Columbia University New York, NY}

\begin{document}

\section{Introduction}

Fixing to the smooth, covariant, Coulomb and Landau gauges can be useful for lattice field theorists. The matching of lattice results to continuum perturbation theory, of particular interest for vertex functions, demands the use of a fixed gauge, usually Landau. Coulomb gauge has other benefits; it can be used for the creation of gauge-fixed wall source fermion propagators or for measurements of the static quark potential.

The Landau and Coulomb gauges are two of the easiest (that have useful continuum analogs) to implement on the lattice as their gauge conditions can be defined as the minimisation of a functional. Such a description allows us to define a smooth deformation of the fields in which we can systematically move toward a local minimum of the functional. This is the aim of many numerical optimisation techniques and several methods are available \cite{Mandula:1987rh,Mandula:1990vs,deForcrand:1989im}.

Conjugate gradient methods were first introduced in \cite{zbMATH03075195} to solve systems of equations and were generalised to the associated problem of solving non-linear, non-quadratic functions in \cite{Polyak196994,Fletcher01011964,Polak1969}. We show how a non-linear conjugate gradient method can be used with Fourier Acceleration to create a very fast gauge fixing procedure for both Coulomb and Landau gauge with exact and approximate definitions of the gauge field. We also comment on the ineffectiveness of fixing to these gauges using a so-called ``optimal'' steepest descent (SD) routine that tunes the step-size to best minimise the functional at each iteration.

\section{Fields and Definitions}

Lattice gauge fields (the $A_\mu$'s) of an $N_d$-dimensional, strongly interacting gauge theory are formally defined by their path-ordered parallel transport matrices (known as lattice links),
\begin{equation}
 U_{\mu}\left(x+a\frac{\hat{\mu}}{2}\right) = P[e^{ig_0 \int_{x}^{x+a\hat\mu} A_\mu( x )dx_\mu }] \approx e^{iag_0A_\mu\left(x+a\frac{\hat\mu}{2}\right)},
\end{equation}
and are well described as living halfway between the sites $x$ and $x+a\hat{\mu}$ of the lattice, where $a$ is the lattice spacing, $g_0$ is the bare coupling and $\mu$ is the field's polarisation direction. The links are elements of the group $\textrm{SU}(N_C)$, the fields are elements of the Lie Algebra $\mathfrak{su}(N_C)$ with $N_C$ being the number of color charges of the theory.

The fields cannot be transformed in a gauge-covariant manner directly in lattice simulations, instead they have to be altered by gauge transformation,
\begin{equation}\label{Eq:gtrans}
 U^{(g)}_\mu\left( x + a \frac{\hat\mu}{2}\right) = g(x)U_\mu\left( x + a \frac{\hat\mu}{2}\right)g(x+a\hat\mu)^{\dagger}.
\end{equation}
Such a transformation preserves the gauge invariance of the action. 

The techniques we consider in this work for fixing the gauge (gradient descent methods) require access to the gauge fields directly. There is some freedom in the choice of field definition: a common one is what we call the ``Hermitian Projection'' and is accurate only up to terms of $O\left(A_{\mu}\left( x+a\frac{\hat\mu}{2}\right)^3\right)$,
\begin{equation}
\begin{aligned}
 ag_0A_{\mu}\left( x+a\frac{\hat\mu}{2}\right)= &\frac{U_{\mu}\left( x+a\frac{\hat\mu}{2}\right)-U_{\mu}\left( x+a\frac{\hat\mu}{2}\right)^{\dagger}}{2i} \\
 &-\frac{1}{2iN_C}\textrm{Tr}\left[U_{\mu}\left( x+a\frac{\hat\mu}{2}\right)-U_{\mu}\left( x+a\frac{\hat\mu}{2}\right)^{\dagger} \right]\cdot I_{N_c\times N_c}.
 \end{aligned}
\end{equation}
Some authors have also investigated the use of exact logarithmic fields \cite{Furui:1998cg,Ilgenfritz:2010gu}. These will be used here to illustrate the general performance properties of our Fourier Accelerated Conjugate Gradient algorithm.

Cayley-Hamilton theorem states that every matrix is a solution of its own characteristic equation. In terms of our links the following finite expansion can be used (where the f's are necessarily complex and $I$ is the identity matrix) \cite{PhysRevD.69.054501},
\begin{equation}
 U_{\mu}\left(x+a\frac{\hat{\mu}}{2}\right) = f_0\cdot I_{N_C\times N_C} + f_1A_\mu\left( x + a\frac{\hat{\mu}}{2}\right) + \cdots +f_{N_C-1}A_\mu\left( x + a\frac{\hat{\mu}}{2}\right)^{N_C-1}.
\end{equation}

The f's are the solution to the following Vandermonde equation,
\begin{equation}\label{chap5:mat:vandermonde}
\begin{pmatrix}
1&q_0&q_0^{2}&\cdots&q_0^{N_C-1} \\
1&q_1&q_1^{2}&\cdots&q_1^{N_C-1} \\
\vdots&\vdots&\vdots&\cdots&\vdots \\
1&q_{N_C-1}&q_{N_C-1}^{2}&\cdots&q_{N_C-1}^{N_C-1}
\end{pmatrix}
\begin{pmatrix}
f_0 \\
f_1 \\
\vdots \\
f_{N_C-1} \end{pmatrix}
=\begin{pmatrix}
 e^{iq_0} \\
 e^{iq_1} \\
 \vdots \\
 e^{iq_{N_C-1}}
 \end{pmatrix}.
\end{equation}
The $q$'s are the Eigenvalues of $A_\mu\left(x+a\frac{ \hat\mu}{2} \right)$ and the $e^{iq}$'s are the Eigenvalues of the link matrix $U_\mu\left(x+a\frac{ \hat\mu}{2} \right)$. For our logarithm, we compute the $e^{iq}$'s from the characteristic equation of the link matrix $U_\mu\left(x+a\frac{ \hat\mu}{2} \right)$ and take their complex argument to define the $q$'s on the principal branch. The solution to general Vandermonde systems can become numerically unstable, but for $\textrm{SU}(3)$ well-behaved functions for the f's in terms of the q's can be used \cite{PhysRevD.69.054501}. This leads to a computationally efficient and numerically stable definition for the principal logarithm of $\textrm{SU}(3)$ matrices,
\begin{equation}\label{eq:logroj}
ag_0A_{\mu}\left( x+a\frac{\hat\mu}{2}\right) = \frac{f_2^*U_{\mu}\left( x+a\frac{\hat\mu}{2}\right)-f_2U^{\dagger}_{\mu}\left( x+a\frac{\hat\mu}{2}\right)-2i\Im{\left(f_0 f_2^*\right)}\cdot I_{3\times3}}{2i\Im{\left(f_1 f_2^*\right)}}.\nonumber
\end{equation}

We define the lattice finite difference of the gauge fields by,
\begin{equation}
 a\Delta_\mu A_\mu(x) = \sum_\mu \left[A_\mu\left( x+ a\frac{\hat\mu}{2} \right) - A_\mu\left( x - a\frac{\hat\mu}{2} \right)\right].
\end{equation}

It will be useful to monitor our algorithm (at it's $n^\textrm{th}$ iteration) by defining the quantity ($V$ is the lattice volume),
\begin{equation}
\Theta^{(n)} = \frac{1}{N_c V}\sum_x \textrm{Tr}\left[ \left( a\Delta_\mu ag_0 A_\mu^{(n)}(x)\right)^2 \right].
\end{equation}

\section{Fourier Accelerated Steepest Descent}

The continuum gauge condition for Landau gauge (and Coulomb with $\mu$ only in the spatial directions) is $\partial_\mu A_\mu(x)=0$, and the lattice analog of this is $a\Delta_\mu ag_0 A_\mu(x) = 0$. This occurs at the minimum of the gauge fixing functional,
\begin{equation}\label{Eq:functional}
 F(U) = \frac{1}{N_C N_d V} \sum_{x,\mu} \textrm{Tr}\left[ A_\mu\left( x + a\frac{\hat\mu}{2} \right)^2\right].
\end{equation}

A general steepest descent step that drives the gauge fields toward a local minimum of the functional (and thus reduces $a\Delta_\mu ag_0A_\mu(x)$ to 0) is,
\begin{equation}
 g(x) = e^{-i\alpha a\Delta_\mu ag_0 A_\mu(x)},
\end{equation}
followed by a gauge transformation of the links (Eq.\ref{Eq:gtrans}). The parameter $\alpha$ is a small, fixed, positive tuning parameter. This procedure is repeated until some convergence criterion is met.

The authors of \cite{PhysRevD.37.1581} suggested that a re-scaling of the Eigenvalues of the Abelian Laplacian operator ($\Delta^{2}A_\mu(x)$) in momentum space was sufficient to reduce some of the critical slowing down in each steepest descent iteration, this is effectively a preconditioner of the gradient. A Fourier Accelerated Steepest Descent (FASD) iteration is,
\begin{equation}
 g(x) = e^{\tilde{F} \;\frac{-i\alpha \hat{p}^2_{\textrm{Max}}}{V\hat{p}^2} \;F a\Delta_\mu a g_0 A_\mu(x) }.
\end{equation}
Followed by a gauge transformation. $p^2_{\text{Max}}$ takes the values $4N_d$ and $4(N_d-1)$ for fixing to Landau and Coulomb gauge respectively. $F$ and $\tilde{F}$ are forward and backward Fast Fourier Transforms (FFTs using FFTW \cite{FFTW05} in our implementation). We use the definition of lattice momentum (the $n_\mu$'s are Fourier modes, $L_\mu$ is the lattice length in the $\mu$ direction),
\begin{equation}
\hat{p}^2 = 2\left( N_d - \sum_\mu \cos\left( \frac{2\pi n_\mu}{L_\mu} \right)\right).
\end{equation}
The rest of this paper will present results from gauge group $\textrm{SU}(3)$, $N_d=4$ dimensional lattice gauge theory. In particular, $\text{N}_\text{f}=2+1$, Iwasaki gauge, DWF configurations are used. We will consider three different lattice volumes; the $32^3\times64$ data has a lattice spacing of $a^{-1}\approx 2.38$ GeV, the $24^3\times64$ and $16^3\times32$ configurations' $a^{-1}\approx 1.79$ GeV \cite{RBC:2014tka}.

\subsection{Optimal FASD}

Many popular techniques for fixing the gauge require some form of tuning to be effective, with FASD and over-relaxation being examples. The tuning parameter can depend on the bare coupling, the gauge action, the field definition and the functional being used. If the tuning parameter is too large the routine will not converge, too small and the routine will have poor performance. The optimal FASD routine, and the Fourier Accelerated Conjugate Gradient algorithm discussed later do not need to be tuned.

We can define a so-called ``optimal'' steepest descent routine by choosing the parameter $\alpha_n$ to be the one which best minimises the gauge functional (Eq.\ref{Eq:functional}) at each iteration of the algorithm. Evaluating the optimal parameter $\alpha_n$ exactly (by e.g. a binary search) would be prohibitively expensive as each evaluation of what we call a probe $\alpha^{\prime}$ requires both a lattice-wide gauge transformation and subsequent functional evaluation. The approach we take instead is to evaluate a selection of probe $\alpha^{\prime}$'s and create a cubic spline interpolation of the probes for which we can solve for the minimum and hence provide a good approximation to the actual minimum. Such a process is called an approximate line search and will be integral to the non-linear conjugate-gradient method discussed in the next section.

\begin{figure}[h!]
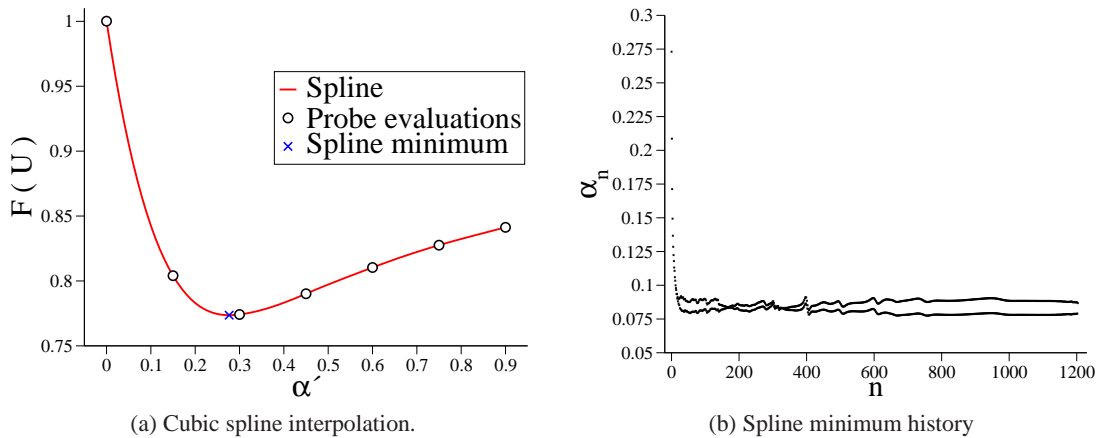

\centering
\subfloat[Cubic spline interpolation.]
  {
  \includegraphics[scale=0.26]{FASD_splines.eps}
  \label{Fig:spline_interp}
  }    
\quad
\subfloat[Spline minimum history]
  {
  \includegraphics[scale=0.26]{FASD_spline.eps}
  \label{Fig:FASD_spline}
  }
\caption[$\:$ ]{Fig.\ref{Fig:spline_interp} shows a cubic spline interpolation and the computation of the spline minimum for a series of probe $\alpha^{\prime}$s. Fig.\ref{Fig:FASD_spline} shows the history of the spline minimum per iteration of the optimal FASD from an example $16^3\times32$ configuration converged to a high accuracy of $\Theta^{(n)}<10^{-14}$.}%
\end{figure}
Fig.\ref{Fig:spline_interp} shows the cubic spline interpolation for a group of probe evaluations for an FASD step. Fig.\ref{Fig:FASD_spline} illustrates the typical history of the cubic spline minimum per iteration of the optimal FASD Landau gauge fixing routine, taken from a single $16^3\times32$ configuration. We see that after the first 50 or so iterations the best $\alpha_n$ is roughly constant at around $0.08$ which is consistent with the best tuning parameter of similar configurations found in the literature \cite{PhysRevD.37.1581,Ilgenfritz:2010gu}. The approximate optimal $\alpha_n$ oscillates each iteration around a central value, illustrating the routine correcting itself from an over or under-estimate of the true optimal $\alpha_n$.

As the optimal FASD shows a long plateau of the best $\alpha_n$ with little variation the cost of the line search is not worthwhile for the FASD algorithm. This is because it is roughly $N\times$ the cost of one fixed-$\alpha$ iteration ($N$ is the number of probe $\alpha^{\prime}$s) and we need at least $N=3$ to compute a cubic spline interpolant. If the number of iterations is not reduced by a factor of $N$ the optimal FASD is less optimal (in the sense of time taken to reach a desired precision) than the fixed-$\alpha$ FASD, which we find to be the case.

\section{Conjugate Directions in Gauge Fixing}

A detailed description of the of the Fourier Accelerated Conjugate Gradient (FACG) algorithm is presented in Alg.1 and Alg.2 of \cite{Hudspith2015115} and will not be reproduced here.

\begin{figure}[ht!]
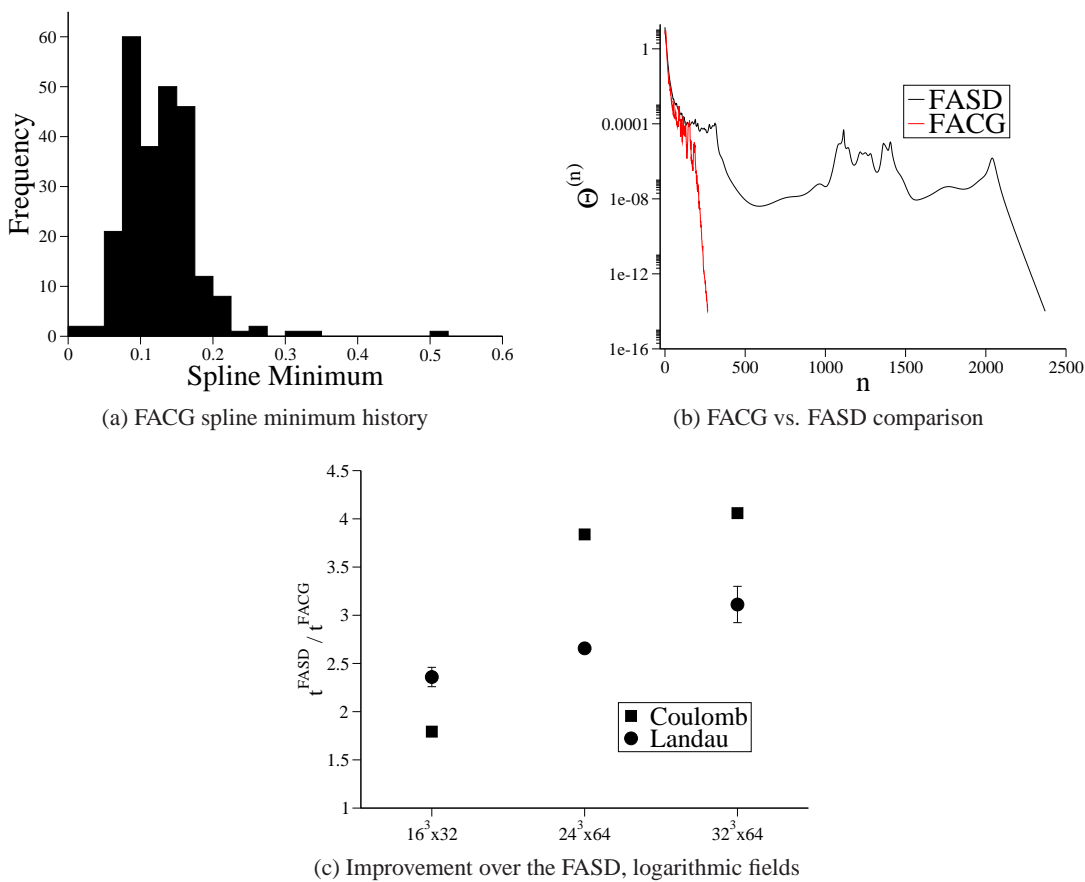

\centering
\subfloat[FACG spline minimum history]
{
\includegraphics[scale=0.25]{FACG_spline.eps}
\label{Fig:spline_hist}
}
\quad
\subfloat[FACG vs. FASD comparison]
{
\includegraphics[scale=0.25]{FACG_v_FASD_acc_iters.eps}
\label{Fig:comparison}
}
\newline
\subfloat[Improvement over the FASD, logarithmic fields]
{	
  \includegraphics[scale=0.26]{scaling_log.eps}
  \label{Fig:log_scalings}
}
\caption[]{Fig.\ref{Fig:spline_hist} shows the optimal $\alpha_n$'s of the FACG algorithm for an example $16^3\times32$ configuration being fixed to $\Theta^{(n)}<10^{-14}$. Fig.\ref{Fig:comparison} shows the number of iterations required for the (fixed-$\alpha$) FASD and FACG algorithms to reach the same accuracy for the same configuration, using the Hermitian Projection definition of the gauge fields. Fig.\ref{Fig:log_scalings} illustrates a measure of the effective speed up of our FACG compared to our FASD routine for the logarithmic field definition for 3 different lattice volumes.}
\end{figure}
  
Fig.\ref{Fig:spline_hist} shows a histogram of $\alpha_n$ taken from the probes $(0.0,0.15,0.3,0.45,0.6)$ for every iteration of the configuration used in the previous section. We can see that most of the optimal $\alpha_n$'s lie in the range $[0,0.3]$ and so we choose the three probe evaluations $(0.0,0.15,0.3)$ for our routine. The probe at $0.0$ is very beneficial in bounding the minimum and is cheap to evaluate as no gauge transformations are needed. If the spline minimum is negative, we set $\alpha_n$ to $0.0$ (i.e. we generate a new descent direction) or if it is greater than $0.3$ we set it to $0.3$ knowing that the algorithm will accommodate for a sub-optimal step. 

We provide a comparison of the FACG algorithm (using the 3 probes) to the fixed-$\alpha$ FASD algorithm in Fig.\ref{Fig:comparison} showing the number of iterations required to reach $\Theta^{(n)}<10^{-14}$ for a single configuration, although this behaviour is ubiquitous. We see that for this configuration there is a factor of around $10$ reduction in the number of iterations to reach a high level of convergence. As mentioned for the optimal FASD routine, if the factor of reduction in the number of iterations is greater than the number of probe evaluations ($N$) then the routine is beneficial. 

In \cite{Hudspith2015115}(Fig.1) we show for the Hermitian Projection definition of the fields a factor of 2 to 4 improvement in the average time taken to converge to a high fixed accuracy was obtained for the FACG routine compared to the FASD. In Fig.\ref{Fig:log_scalings}, we show that such speedup is also available for the logarithmic field definition for the three volumes considered. The measure of improvement was taken to be the time taken for 25 randomly gauge transformed copies of the same configuration at each volume for the FASD algorithm divided by the time taken for the FACG algorithm.

\section{Conclusions}

We have presented results using the FACG algorithm for fixing to Landau and Coulomb gauge. It is shown to be faster than our implementation of the FASD algorithm for both the logarithmic and Hermitian Projection field definitions. We have illustrated why an optimal FASD algorithm does not provide a faster method than the fixed-$\alpha$ procedure, and have shown why the literature value of $\alpha=0.08$ is often a good choice for the tuning parameter, although the exact optimal tuning parameter is dependent on several factors. Unlike the fixed-$\alpha$ FASD or over-relaxation methods, the FACG does not require one to tune a parameter.

The FACG algorithm converges in fewer iterations than the FASD, and requires only one set of FFTs per iteration; if FFTs are particularly costly this method will show greater improvement over the FASD than presented here. The FACG method (as it is a gradient descent method) allows for options where standard relaxation techniques cannot be used; different gauge field definitions or alternate functionals can be accommodated easily.

\section{Acknowledgements}

The results produced in this work were generated on Columbia University's cluster ``CUTH'' using the package GLU (\href{https://github.com/RJhudspith/GLU}{https://github.com/RJHudspith/GLU}). RJH would like to thank R. Lewis. RJH is supported by the Natural Sciences and Engineering Research Council of Canada.

\bibliographystyle{JHEP}
\bibliography{Lattice2014.bib}

\end{document}